\documentclass{aa}

\usepackage{graphics}
\usepackage{rotate}

\newcommand{\Msun}{{\rm M}_{\odot}}

\begin{document}

\thesaurus{02         
           (11.03.1; 12.03.3; 12.03.4)}  

\title{Quantifying Substructure in Galaxy Clusters}

\titlerunning{Quantifying Substructure in Clusters}

\author{Alexander Knebe \inst{1,2}
        \and
        Volker M\"uller \inst{1}
        }

\offprints{A. Knebe}

\institute{$^1$Astrophysikalisches Institut Potsdam (AIP), 
           An der Sternwarte 16, 14482 Potsdam, Germany\\
           $^2$Theoretical Physics, 1 Keble Road, Oxford OX1 3NP, England }

\date{Received ; accepted}

\maketitle

\begin{abstract}

Substructure in galaxy clusters can be quantified with the robust $\Delta$
statistics (Dressler and Shectman 1988) which uses velocity kinematics and sky
projected positions.  We test its sensitivity using dissipationless numerical
simulations of cluster formation.  As in recent observations, about 30\% of the
simulated clusters show substructure, but the exact percentage depends on the
chosen limit for defining substructure, and a better discriminator is the
distribution function of the $\Delta$ statistics.  The Dressler-Shectman
statistics correlate well with other subcluster indicators, but with large
scatter due to its sensitivity to small infalling groups and projection effects.

\keywords{galaxies: clusters; cosmology -- observations, theory}

\end{abstract}


\section{Introduction}

During the last two decades, substructure was detected in a significant fraction
of groups and clusters of galaxies, both as double or secondary maxima in the
sky-projected galaxy distribution of clusters (cf.  \cite{baier} and Geller \&
Beers 1982), and as deviations of spherical symmetry in X-ray contour maps
(\cite{forman}).  However, quantifying substructure in clusters is a non-trivial
problem.  Analyzing substructure indicators with numerical simulations, West et
al.  (1988) concluded that many subclumps may be pure chance projections, and
that the remaining abundance of substructure can barely discriminate between a
wide range of cosmological scenarios.  Projection effects are less pronounced in
X-rays, and recent X-ray data have provided more reliable information on the
abundance of substructure (cf.  \cite{mohr93} and Buote \& Xu 1997).

In the optical spectral range, one gets better results by including velocity
information for a large sample of cluster galaxies.  Substructure was identified
using subgroup velocity statistics in clusters (\cite{dressler}), finding
velocity offsets of cD galaxies with respect to the major cluster (\cite{bird}),
employing the hierarchical tree algorithm (\cite{serna}) and wavelet analysis
(\cite{girardi}).  Recently, Dressler \& Shectman's statistics were used by
Zabludoff \& Mulchaey (1998) for 6 poor groups, and by Solanes et al.  (1998)
for 67 rich clusters from the ENACS survey.  These different analyses agreed in
$(30 - 40)\%$ of clusters which showed statistically significant substructure.
But it appears that this amount depends on the analysis method (\cite{pinkney})
and on the imposed reliability criterion.

The existence of substructure in clusters suggests that the clusters are young
objects since loosely bound subgroups can survive only a few cluster crossing
times, i.e., shorter than the Hubble time.  In the framework of the hierarchical
structure formation scenario, clusters grow mainly by mergers of smaller objects
and by accretion; substructure point to recent major mergers.  Starting from
these ideas and the early termination of growth of density perturbations in a
low density universe, the abundance of substructure was suggested as an
effective measure of the mean matter density in the universe (Richstone, Loeb \&
Turner 1992, Bartelmann , Ehlers \& Schneider 1993, Kauffman \& White 1993, and
Lacey \& Cole 1994).  The interpretation of subclusters as recent mergers might
explain also some of the properties of cD cluster galaxies, such as their
orientation with respect to the environment (\cite{west94}) and the peculiar
velocity distribution in rich clusters (\cite{merritt}).

The velocity kinematics as a signature of substructure traces the galaxy
distribution, therefore we employed dissipationless simulations in a large
cosmological environment.  The cosmological models take COBE-normalized
perturbation spectra in four cosmological scenarios.  Earlier theoretical
studies concentrated on the analysis of the density contrast of clusters and the
density profile (\cite{jing}, \cite{thomas}).  There, relatively small
differences in the cluster properties are found when comparing cluster profiles
at the same overdensity.  This concerns mainly the properties of relaxed
clusters.  Here we study in particular clusters which represent deviations from
an equilibrium.  For quantifying substructure in X-ray profiles of clusters,
hydrodynamic simulations of cluster formation have to be employed, and early
attempts at this have already yielded first promising results (\cite{cen} and
Valdarnini, Ghizzardi \& Bonometto 1999).

The outline of the paper is as follows.  First we analyze the method for
identification of substructure in galaxy clusters selected from numerical
simulations.  Then we apply the algorithm to different cosmological models. 
In Section 4 we compare the substructure measure with other methods and 
with observational data.  We conclude with a discussion of our results.


\section{Method}

The Dressler \& Shectman (1998) statistics evaluate the velocity kinematics of
galaxy groups identified in sky projected clusters.  To be specific, one takes a
number of neighbours $N_{\rm nn}$ from each galaxy in the projection, determines
the mean velocity $\overline{v}_{\rm local}$ and velocity dispersion $\sigma_{v,
\rm local}$ of the subsample, and compares this with the mean velocity
$\overline{v}$ and velocity dispersion $\sigma_{v}$ of the whole group,

\begin{equation} 
\delta_i^2 = \displaystyle \frac{N_{\rm nn}}{\sigma_v^2}
\left[(\overline{v}_{\rm local} - \overline{v})^2 + (\sigma_{v, \rm
local} -\sigma_v)^2 .  \right]
\end{equation} 

\begin{figure}
      \resizebox{\hsize}{!}{\includegraphics{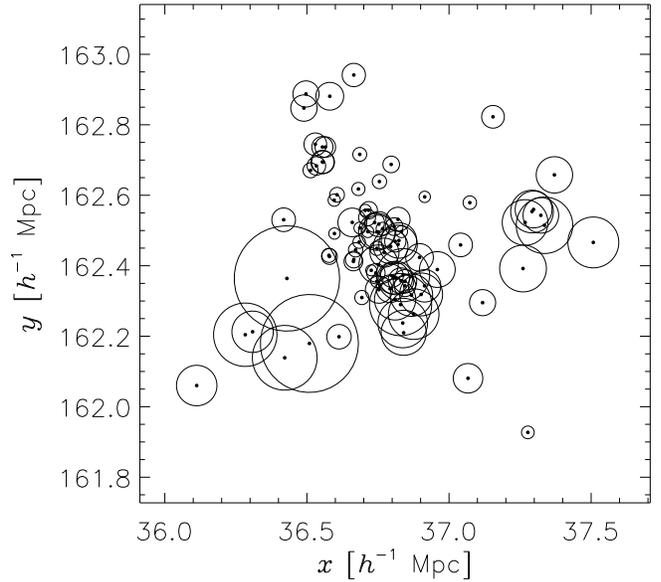}}
      \caption{Bubble plot for a particle group containing substructure.}
      \label{bubble}
    \end{figure}

A measure of the amount of clumpiness in the cluster is the sum of the
individual positive $\delta_i$ over all cluster galaxies $N$,
\begin{equation} 
\Delta = \sum_{i=1}^N \delta_i , 
\end{equation}
which is called the delta-deviation.  The $\Delta$-deviation is large for groups
with kinematically distinct subgroups.  Fig.~\ref{bubble} illustrates the
statistics for a simulated typical dark matter cluster selected with a linking
length of 0.2 times the mean interparticle spacing.  Around each point a circle
is plotted with radius proportional to $\exp(\delta_i)$.  The cluster has a
pronounced centre indicated by the central particles in the figure decorated
with the small circles.  To the right and above from the centre there are some
subclumps which do not represent especially important subgroups with decoupled
kinematics, but most probably they are small satellites slowly falling onto the
group centre.  To the left, there are some particles with large $\delta_i$ that
contribute mostly to the cumulative $\Delta$-deviation.  The cluster in
Fig.~\ref{bubble} represents an example for marginally reliable substructure, 
we base our later statistical analysis on clusters with more pronounced
substructure.  An obvious advantage of the method is that no {\it a priori}
selections or assumptions about the positions of subclumps have to be imposed.


In order to test and calibrate the statistics, we have compared five clusters of
masses of about $10^{15} h^{-1}\Msun$ (1000 particles), $5 \cdot 10^{14}
h^{-1}\Msun$ (500 particles), and $2 \cdot 10^{14} h^{-1}\Msun$ (200 particles)
in a standard CDM simulation.  For each mass we study clusters both with and
without obvious substructure as decided by eye from the 3-dimensional
distribution.

\begin{figure}
      \resizebox{\hsize}{!}{\includegraphics{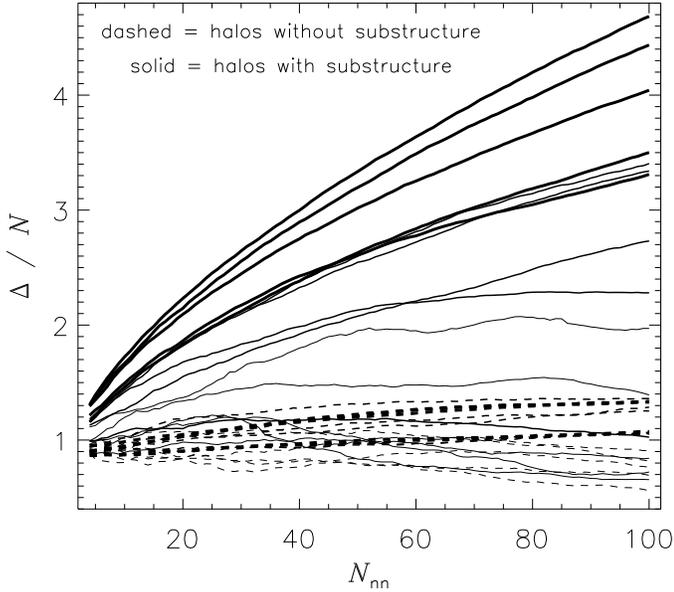}}
      \caption{Variation of $\Delta/N$ with the number of neighbours 
      $N_{\rm nn}$. The thickness of lines is proportional to the mass.}
      \label{nlocal}
    \end{figure}

The $\Delta$-deviation statistics depend on the number of cluster galaxies with
measured redshift $N$ that can be included in the analysis, and on the number of
neighbours in tested subgroups $N_{nn}$.  If one assumes a random distribution of
velocities one expects values $\Delta \propto N$ (Dressler \& Shectman 1988),
and for galaxy clusters with substructure, $\Delta \geq N$ (\cite{pinkney}).
Therefore we normalize $\Delta$ with the total number of galaxies investigated, 
and in Fig.~\ref{nlocal} we plotted $\Delta / N$ against the number of neighbours
$N_{\rm nn}$ in subgroups.  The selected groups with substructures (solid lines)
show increasing values if the clusters have a large mass.  Only the poor
clusters lead to a poor discrimination of substructure.  The curves suggest the
use of a rather high value for the number of neighbours, while Pinkney et al.
(1996) propose $N_{\rm nn} = N^{1/2}$.  The maximum for low mass objects (more
than 200 particles) lies at about 25 neighbours which we therefore took as a
reasonable value for the number of nearest neighbours in the following analysis.
We took a fixed number in order to be independent of the sampling rate of
velocities.  A similar plot of $\Delta / N$ over the number of objects in
clusters randomly selected demonstrates a very stable discrimination of clusters
with substructure from those that are in equilibrium, and $\Delta / N$ remains
almost constant for values $N \ge 100$.  Ideally this would be the number of
required redshifts to be measured for analyzing clusters for substructure.

\section{Application to Cosmological Simulations}

We apply the substructure statistics to a set of cosmological models that are
currently under investigation.  Three of the models are COBE normalized
according to the prescription of Bunn \& White (1997) where we assume pure
adiabatic perturbations and a baryon content of $\Omega_b h^2 = 1.3 \cdot
10^{-3}$.  The standard CDM model (SCDM1) is taken as a reference model despite
its enhanced power at cluster scales.  In particular it produces too high a
cluster abundance.  As more realistic alternatives we take an OCDM model with
$\Omega_0 = 0.5$ and a $\Lambda$CDM model with $\Omega_0 = 0.3$ and a
cosmological constant to provide spatial flatness (i.e.  $\Omega_{\Lambda,0} =
0.7$).  Both models are promising since they lead nearly to the observed cluster
abundance (\cite{eke}), and they represent models which reproduce the observed
superclustering of galaxies (\cite{doroshkevich}).  In addition to these three
models we have used the output of the SCDM1 model at a redshift $z=1.0$ to
ensure the correct cluster abundance (SCDM2).  The model parameters are
summarized in Table~\ref{simuparam}.  In particular it is shown that the
simulations are running in boxes that are identical in physical size $L$,
thereby making small differences in the mass resolution $m_p$ despite the
different density parameters.  All models were run using the adaptive P$^3$M
code of Couchman (1991) with $128^3$ particles.  An analysis of the
virialization of clusters in these simulations can be found in a previous paper
(Knebe \& M\"uller 1999).

\begin{table}
\caption{Cosmological scenarios and percentage of substructure. 
         The box size~$L$ is given in~$h^{-1}$Mpc, and the particle 
         mass~$m_p$ in units of 
         $10^{11} h^{-1}{\rm M}_{\odot}$.}
\label{simuparam}
 \begin{tabular}{|l||c|c|c|c|c|c|c|} \hline
                              & $\Omega_0$ 
                              & $h$ 
                              & $\sigma_8$   & $L$  & $m_p$ 
                              & $P(\Delta/N \geq 1.4)$ \\ \hline \hline
 {\bf SCDM1}                  & 1.0 & 0.5 & 1.18 & 200 & 11 & 33\% \\ \hline
 {\bf SCDM2}                  & 1.0 & 0.5 & 0.53 & 200 & 11 & 43\% \\ \hline
 {$\bf \Lambda{\rm \bf CDM}$} & 0.3 & 0.7 & 1.00 & 280 & 9  & 27\% \\ \hline
 {\bf OCDM}                   & 0.5 & 0.7 & 0.96 & 280 & 15 & 30\% \\ \hline
 \end{tabular}
\end{table}

\begin{figure}
      \resizebox{\hsize}{!}{\includegraphics{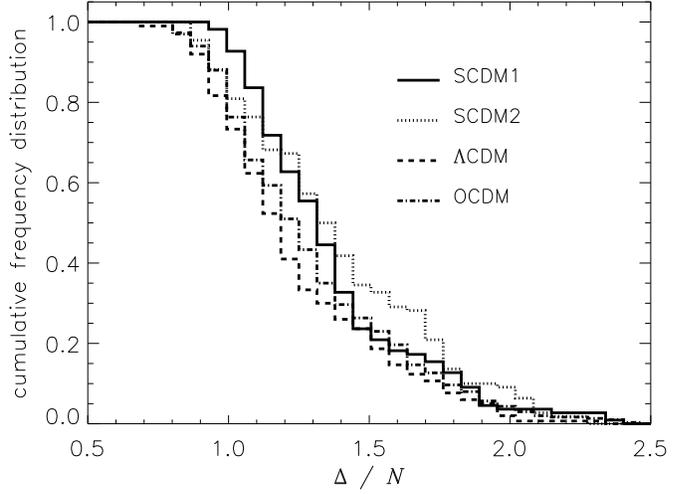}}
      \caption{Cumulative probability distribution of $\Delta$ for different
               cosmological models.}
      \label{Dmodel}
    \end{figure}

As explained in the introduction, we expect differences in the abundance of
substructure for different cosmologies, and therefore in the probability
distribution of the delta-deviation.  In a low-density universe (open or flat),
structure formation ceases at earlier times compared to the SCDM models.  This
means that clusters in low-density universes should show less substructure since
they formed earlier and therefore had more time to virialize.  We always
obtained strongly varying delta-deviations, therefore, we only analyze its
abundance distribution.  In Fig.~\ref{Dmodel} we show the cumulative probability
distributions of the delta-deviations $\Delta/N$ of friends-of-friends clusters
selected with dimensionless linking length (in terms of the mean interparticle
separation) $ll = 0.2$, 0.17, and 0.16 for SCDM, OCDM, and {$\Lambda$}CDM,
respectively.  We show these distributions for a fixed number density of
simulated clusters $n_{cl} = 10^{-5} h^3\rm{Mpc}^{-3}$.

The distributions show differences which underline the cited expectation.  In
particular, both SCDM simulations have higher probabilities of substructure
than the low density models.  Especially the SCDM2 model shows a large number of
clusters with $\Delta/N \ge 1.4$.  On the other hand, the low density models
$\Lambda$CDM and OCDM have a similar distribution of the delta-deviation, with
$\Lambda$CDM lying below the OCDM model in agreement with the expectation due to
the lower density parameter.  To quantify the results, the probability of
substructure is given in Table~\ref{simuparam} as the percentage of clusters
with $\Delta/N > 1.4$.  We find that about $30\%$ of the investigated objects
show a significant level of substructure.  This coincides with the observed
probability of substructure in recent studies using the delta-deviation
(\cite{dressler}, Zabludoff \& Mulchaey 1998, Solanes, Salvado-Sol'e \&
Honz'alez-Casado 1999).  To check the significance, a Kolmogorov-Smirnov test
was employed that is summarized in Table~\ref{KS}.  The first column gives the
maximum distance of the compared distributions, whereas the second column is the
probability (in percent) that the models can be distinguished by the abundance
of substructure as quantified by the delta-deviation (100~\% means `different' and
0~\% means `identical').  The values in Table~\ref{KS} show that the probability
distributions for the SCDM models differ from that for the {$\Lambda$}CDM and
OCDM models.  On the other hand, the low density models with and without a
cosmological constant are very similar, i.e., they cannot be discriminated.  For
all four models, the abundance of substructure depends more on the imposed limit
in the delta-deviation than on the difference between models.  Some degree of
substructure is typical for all simulated clusters.

\begin{table}
\caption{Kolmogorov-Smirnov (KS) test for the cumulative distributions.
         First column for each redshift gives maximum distance $D$,
         and second column the significance level.}
\label{KS}
 \begin{tabular}{|l||c|c|} \hline
    Simulations               & difference & reliability \\ \hline \hline
  \raisebox{4mm}{}{\bf SCDM1} vs. {\bf SCDM2}
                              &  0.14 & 89 \%                     \\ \hline
  \raisebox{4mm}{}{\bf SCDM1} vs. {$\bf \Lambda{\rm \bf CDM}$}
                              &  0.24 & 96 \%                     \\ \hline
  \raisebox{4mm}{}{\bf SCDM1} vs.  {\bf OCDM}    
                              &  0.18 & 95 \%                     \\ \hline \hline
  \raisebox{4mm}{}{\bf SCDM2} vs. {$\bf \Lambda{\rm \bf CDM}$}     
                              &  0.27 & 99 \%                     \\ \hline
  \raisebox{4mm}{}{\bf SCDM2} vs.  {\bf OCDM} 
                              &  0.17 & 97 \%                     \\ \hline \hline
  \raisebox{4mm}{}{$\bf \Lambda{\rm \bf CDM}$}  vs.  {\bf OCDM}
                              &  0.11 & 48 \%                     \\ \hline
 \end{tabular}
\end{table}

\begin{figure}
      \resizebox{\hsize}{!}{\includegraphics{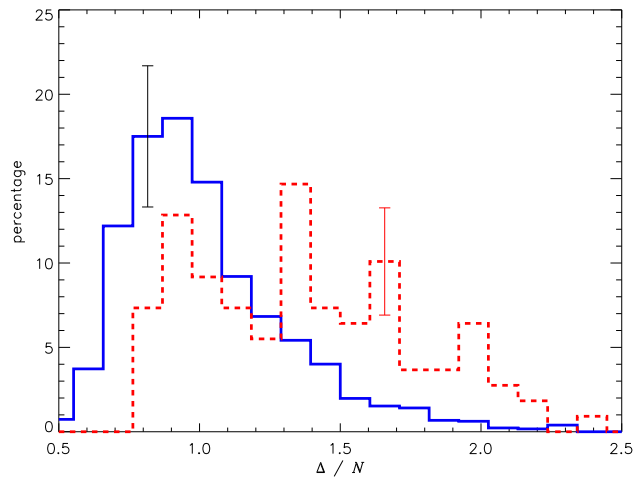}}
      \caption{Probability distribution of $\Delta$ for the SCDM1
               model distinguishing virialized (solid)
               and unvirialized (dashed) particle groups (with Poisson 
               error bars).}
      \label{Dbound}
    \end{figure}

As we have recently shown in simulations (Knebe \& M\"uller 1999), unvirialized
particle groups can be identified with recent or ongoing mergers which could be
a possible explanation for substructure.  For this reason we have separated
`virialized' and `unvirialized' clusters.  The corresponding differential
distribution of the delta-deviation in the SCDM1 model is shown in
Fig.~\ref{Dbound}.  Unvirialized clusters clearly have more substructure than
relaxed systems.  Substructure is therefore a clear sign of incomplete
relaxation.  Note that the distributions of virialized and unvirialized groups
are separately normalized, but the number of unvirialized groups is always much
smaller.  The same behaviour is also found for the other models.

\section{Comparison with Other Indicators of Substructure}

For testing the delta-deviation against other substructure indicators, we have
also employed the friends-of-friends algorithm with a smaller linking length.
Similar to the hierarchical tree (\cite{serna}, Klypin et al.  1999), we
determine the mass fraction of the two most massive subgroups to produce 
multiplicity statistics.  To this aim, we build FOF clusters with half of the
linking length of the original analysis.  Then the total mass $M$ of each
particle group of the first analysis is decomposed into
\begin{equation} M = m_1 + m_2 + m_{\rm rem} \end{equation} 
where $m_1$ and $m_2$ are the masses of the two most massive subclumps.  The
remaining mass $m_{\rm rem}$ consists of all the constituents lying in low
density parts of the cluster, probably containing mainly satellites recently
accreted. We define the  multiplicity of each cluster to be
\begin{equation} 
M_p = \displaystyle \frac{m_1 + m_2}{m_1}.
\end{equation} 
For $M_p \approx 1$ we do not resolve substructure because the second most
massive subgroup does not contribute to the total mass of the object.  But for
$M_p \approx 2$ both subclumps are of comparable mass and therefore we have a
cluster with a possible double structure.  A probable origin is a big merger of
almost equal-mass progenitors.

\begin{figure}
      \resizebox{\hsize}{!}{\includegraphics{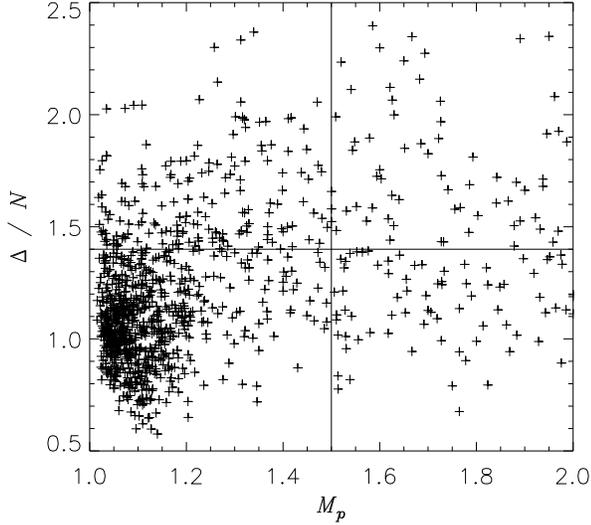}}
      \caption{Correlation for $\Delta/N$ vs. multiplicity $M_p$ in the 
               SCDM1 model. The lines $\Delta/N=1.4$ and $M_p=1.5$ show 
               the adopted criteria for substructure.}
      \label{multidelta}
    \end{figure}

Fig.~\ref{multidelta} shows the correlation of $\Delta/N$ and $M_p$ for the
SCDM1 simulation.  The other models look similar.  The lines separate regions
where both statistics indicate there to be substructure.  One notes a weak
correlation with a wide scatter and a large number of outliers.  The points in
the upper left part show high values of the delta-deviation for which the
multiplicity test shows that the subgroups are not very massive.  So, they are
probably small satellites falling onto the main cluster as seen, e.g., in
Fig.~\ref{bubble}.  The lower right part shows clusters with subclumps of nearly
equal masses, but no velocity deviations.  A possible explanation are projection
effects.

\begin{figure}
      \resizebox{\hsize}{!}{\includegraphics{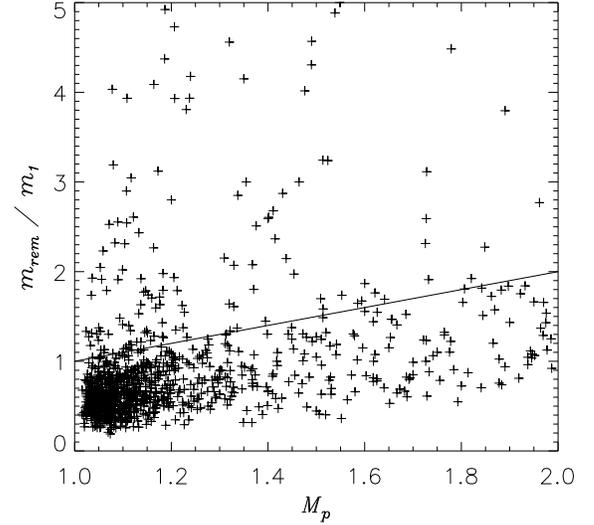}}
      \caption{Influence of the rest mass $m_{\rm rest}$ on the 
               multiplicity $M_p$ in the SCDM1 model. The straight line gives 
               the equality between the masses of two biggest clumps and the 
               rest.}
      \label{multimass}
    \end{figure}

Another interesting point is the relationship of the remaining mass $m_{\rm
rem}$ to the mass of the two most massive subclumps $m_1$ and $m_2$.  If $m_{\rm
rem}$ is not negligible, we expect a loosely bound cluster with a large portion
of accreted material which may be formed recently.  In Fig.~\ref{multimass} we
show the correlation between $m_{\rm rest}/m_1$ and the multiplicity $M_p$ for
the SCDM1 simulation.  The straight line gives the equality between $m_1+m_2$
and $m_{rem}$.  Above it we find loosely bound clusters which are obviously
exceptions (about 13\%).  Below this line, the influence of recently accreted
material is small.  The loosely bound clusters lie mainly in the range where the
multiplicity distribution indicates no significant substructure.  This means
that a steady accretion of material onto galaxy clusters tends to destroy
substructure.  Also we checked that a large accretion leads mostly to low values
of $\Delta / N$.

\begin{figure}
      \resizebox{\hsize}{!}{\includegraphics{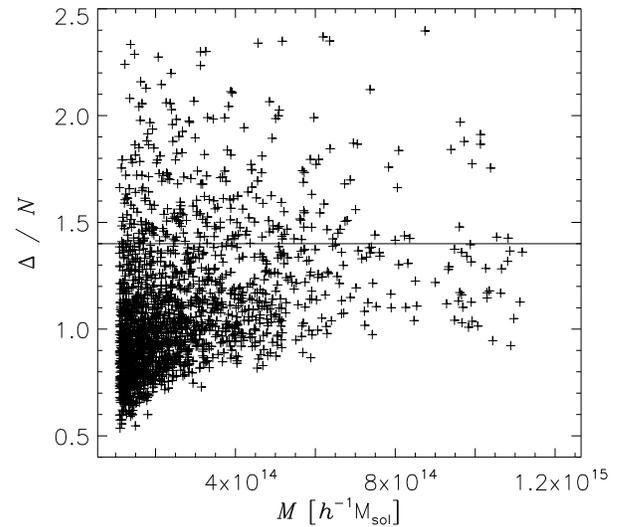}}
      \caption{Substructure dependence on mass of the galaxy cluster in the 
              SCDM1 model, where the line $\Delta/N=1.4$ gives the adopted 
              substructure criterion.}
      \label{Dmodelmass}
    \end{figure}

In Fig.~\ref{Dmodelmass} we show the dependence of identified substructure on the
cluster mass $M$ in the SCDM1 model.  We find particle groups with and without
substructure over the whole mass range, but the scatter for light clusters is
much larger.  When restricting to higher mass objects, the percentage of groups
with substructure increases, and it leads to a value of about 30\%, as in
observed cluster samples (note that above we took a fixed cluster density which 
are taken from the same high mass range). 

\section{Conclusions}

We quantified substructure in simulated clusters of galaxies using the
delta-deviation statistics proposed by Dressler \& Shectman (1988).  Towards this
aim, we studied the scaling of the statistics with the multiplicity of the total
cluster and with the multiplicity of subgroups.  Even with an optimal choice of
both parameters which requires a large numbers of observed redshifts, we get a
broad distribution for the statistics.  The amount of subclustering depends on
the chosen criterion, and we can reasonably reproduce the amount of substructure
found recently in galaxy clusters redshift surveys.  The chosen criterion
($\Delta/N > 1.4$) is similar to that used in observations, but the scatter is
high, and reasonable changes may lead to substructure in the range of $(20 -
50)\%$.  The difference between different cosmologies is significant and a
measure of the mean matter density of the universe.  Among the models studied,
$\Lambda$CDM shows the smallest percentage, and SCDM2 the highest percentage of
subclustering.  We recommend the cumulative distribution of the delta-deviation
as a natural quantifier of substructure.  The differences in its distribution
requires large catalogues of galaxy clusters, each with about 50 redshift
measurements.  On the other hand, it is shown that subclustering is a typical
property of cluster formation in hierarchical theories of structure formation.
In recent hydrodynamical simulations, a high percentage of substructure (in 4 of
10 clusters) were found in a low density $\Lambda$CDM model (\cite{eke98}).
Therefore, quantifying substructure with a distribution function of a
substructure indicator, as done here with dark matter simulations, seems to be a
prerequisite for discriminating cosmological models.  We suspect this is a
general characteristic for different substructure indicators.

We compared the velocity kinematics as an indicator for subclusters with the
subgroups found as friends-of-friends groups with half of the linking length.
There is a weak correlation, i.e., both statistics define similar structures,
but there is large scatter.  On the one hand, projection effects influence the
delta-deviation, and on the other hand, high values of the delta-deviation can
be produced by small groups with separate velocity kinematics.  These effects
are often stronger than the differences between different cosmological
scenarios.  This may be the reason for the large differences in the literature
on the amount of substructure in galaxy clusters.

It was shown that the delta-deviation is most sensitive to recent big mergers,
and that big mergers are more typical for the high mass clusters.  Furthermore,
they occur more often in high-density than in low-density models.


\begin{acknowledgements}

We acknowledge the use of Couchman's AP3M code in this study.

\end{acknowledgements}


\end{document}